\documentclass[]{aipproc}

\usepackage{rotate}

\usepackage{epsf}

\layoutstyle{6x9}

\SetInternalRegister\hbadness{8000} 

\begin{document}


\begin{titlepage}
\begin{flushright}
\begin{tabular}{l}
DESY 01--162\\
hep--ph/0110278\\
October 2001
\end{tabular}
\end{flushright}

\vspace*{1.7truecm}

\begin{center}
\section{Extraction of $\gamma$}

\vspace*{2.1cm}

\subsection{{\sc Robert Fleischer}}

\vspace*{0.1cm} 

{\it Deutsches Elektronen-Synchrotron DESY, 
Notkestr.\ 85,\\ 
D--22607 Hamburg, Germany}

\vspace{1.5truecm}

\subsection{Abstract}

\vspace*{0.1cm} 

\parbox[t]{\textwidth}{
After a brief look at the well-known standard approaches to determine 
the angle $\gamma$ of the unitarity triangle, we focus on two kinds of 
strategies, employing $B\to\pi K$ modes, and $U$-spin-related $B$ decays. 
Interesting ``puzzles'', which may already be indicated by the present 
$B$-factory data, are pointed out, and the importance of the extraction 
of hadronic parameters, which are provided by these strategies as 
by-products, is emphasized.}

\vspace{2.4cm}
 
{\sl Invited talk given at the\\ 
9th International Symposium on Heavy Flavor Physics,\\ 
Caltech, Pasadena, CA, 10--13 September 2001\\
To appear in the Proceedings}

\end{center}

\end{titlepage}
 
\thispagestyle{empty}
\vbox{}
\newpage
 
\setcounter{page}{1}
 

\title{Extraction of $\gamma$
}

\classification{classification}

\author{Robert Fleischer}{
address={Deutsches Elektronen-Synchrotron DESY,
Notkestr.\ 85, D--22607 Hamburg, Germany},
email={Robert.Fleischer@desy.de}
}

\copyrightyear  {2001}

\begin{abstract}
After a brief look at the well-known standard approaches to determine 
the angle $\gamma$ of the unitarity triangle, we focus on two kinds of 
strategies, employing $B\to\pi K$ modes, and $U$-spin-related $B$ decays. 
Interesting ``puzzles'', which may already be indicated by the present 
$B$-factory data, are pointed out, and the importance of the extraction 
of hadronic parameters, which are provided by these strategies as 
by-products, is emphasized.
\end{abstract}

\date{\today}

\maketitle

\section{Setting the Stage}
The recent observation of CP violation in the $B$ system by the 
BaBar and Belle collaborations \cite{CP-obs} manifests the beginning
of a new era in the exploration of particle physics. One of the
central goals of the $B$-factories is to overconstrain the unitarity 
triangle of the Cabibbo--Kobayashi--Maskawa (CKM) matrix as much as 
possible through independent measurements both of its sides and of 
its angles $\alpha$, $\beta$ and $\gamma$ \cite{CP-revs}. 

A particularly important element in this stringent test of the 
Kobayashi--Maskawa mechanism of CP violation is the direct 
determination of the angle $\gamma$. The corresponding experimental
values may well be in conflict with the indirect results provided by 
the fits of the unitarity triangle, yielding at present 
$\gamma\sim60^\circ$ \cite{UT-fits}. Moreover, we may encounter 
discrepancies between various different approaches to determine 
$\gamma$ directly. In such exciting cases, the data may shed light 
on the physics lying beyond the Standard Model.

\subsection{Key Problem in the Determination of $\gamma$}
At leading order of the well-known Wolfenstein expansion of the 
CKM matrix, all matrix elements are real, apart from
\begin{equation}
V_{td}=|V_{td}|e^{-i\beta} \quad\mbox{and}\quad V_{ub}=|V_{ub}|e^{-i\gamma}.
\end{equation} 
In non-leptonic $B$-meson decays, we may obtain sensitivity on $\gamma$
through interference effects between different CKM amplitudes. Making
use of the unitarity of the CKM matrix, it can be shown that at most
two different weak amplitudes contribute to a given non-leptonic decay
$B\to f$, so that we may write
\begin{equation}
A(B\to f)=|A_1|e^{i\delta_1}+
e^{+i\gamma}|A_2|e^{i\delta_2},
\quad
A(\overline{B}\to\overline{f})=|A_1|e^{i\delta_1}+
e^{-i\gamma}|A_2|e^{i\delta_2},
\end{equation}
where $\gamma$ enters through $V_{ub}$ and the $|A_{1,2}|e^{i\delta_{1,2}}$
denote CP-conserving strong amplitudes. Consequently, the corresponding 
direct CP asymmetry takes the following form:
\begin{equation}
{\cal A}_{\rm CP} 
=\frac{|A(B\to f)|^2-|A(\overline{B}\to \overline{f})|^2}{|A(B\to f)|^2+
|A(\overline{B}\to \overline{f})|^2}
=\frac{2|A_1||A_2|\sin(\delta_1-\delta_2)
\sin\gamma}{|A_1|^2+2|A_1||A_2|\cos(\delta_1-\delta_2)
\cos\gamma+|A_2|^2}.
\end{equation}
Measuring such a CP asymmetry, we may in principle extract $\gamma$. 
However, due to hadronic uncertainties, which affect the strong amplitudes
\begin{equation}
|A|e^{i\delta}\sim\sum_k
\underbrace{C_{k}(\mu)}_{\mbox{pert.\ QCD}} 
\times\,\,\, \underbrace{\langle\overline{f}|Q_{k}(\mu)
|\overline{B}\rangle}_{\mbox{``unknown''}},
\end{equation}
a reliable determination of $\gamma$ is actually a challenge.

\subsection{Major Approaches to Determine $\gamma$}
In order to deal with the problems arising from the hadronic matrix
elements, we may employ one of the following approaches:
\begin{itemize}
\item The most obvious one -- and, unfortunately, also the most challenging 
from a theoretical point of view -- is to try to calculate 
the $\langle\overline{f}|Q_{k}(\mu)|\overline{B}\rangle$.
In this context, interesting progress has recently been made through
the development of the ``QCD factorization'' \cite{QCD-fact,BBNS3} and
perturbative hard-scattering (or ``PQCD'') \cite{PQCD} approaches,
as discussed by Neubert and Keum, respectively, at this symposium. 
As far as the determination of $\gamma$ is concerned, $B\to\pi K$, 
$\pi\pi$ modes play a key r\^ole. 
\item Another avenue we may follow is to use decays of neutral 
$B_d$- or $B_s$-mesons, where interference effects between 
$B^0_q$--$\overline{B^0_q}$ mixing ($q\in\{d,s\}$) and decay processes 
arise. There are fortunate cases, where hadronic matrix elements cancel:
\begin{itemize}
\item Decays of the kind $B_d\to D^{(\ast)\pm}\pi^\mp$, allowing a clean
extraction of $\phi_d+\gamma$ \cite{Bd-strat}, where the 
$B^0_d$--$\overline{B^0_d}$ mixing phase $\phi_d=2\beta$ can 
be fixed through $B_d\to J/\psi K_{\rm S}$. 
\item Decays of the kind $B_s\to D_s^{(\ast)\pm}K^{(\ast)\mp}$, 
allowing a clean extraction of $\phi_s+\gamma$ \cite{Bs-strat}, where 
the $B^0_s$--$\overline{B^0_s}$ mixing phase $\phi_s$ is negligibly 
small in the Standard Model, and can be probed through $B_s\to J/\psi \phi$ 
modes. 
\end{itemize}
\item An important tool to eliminate hadronic uncertainties in the 
extraction of $\gamma$ is also provided by certain amplitude relations.
There are two kinds of such relations:
\begin{itemize}
\item Exact relations between $B^\pm\to K^\pm\{D^0,\overline{D^0},D^0_\pm\}$ 
amplitudes \cite{exact-rel}, where $D^0_\pm$ denotes the CP eigenstates 
of the neutral $D$ system. This approach is realized in an ideal way in 
the $B^\pm_c\to D_s^\pm\{D^0,\overline{D^0},D^0_\pm\}$ system \cite{Bc}. 
Unfortunately, $B_c$-mesons are not as accessible as ``ordinary'' 
$B_u$- or $B_d$-mesons.
\item Amplitude relations, which are implied by the flavor symmetries
of strong interactions, i.e.\ isospin or $SU(3)$ \cite{GRL,SU3-rel}.
In the corresponding strategies to determine $\gamma$, we have to deal 
with $B_{(s)}\to\pi\pi,\pi K,KK$ modes. 
\end{itemize} 
\end{itemize}
In the following discussion, we shall focus on the latter kind of strategies, 
involving the $B\to\pi K$ system, and the $U$-spin-related\footnote{$U$ spin 
is an $SU(2)$ subgroup of $SU(3)_{\rm F}$, relating down and strange 
quarks to each other.} decays $B_d\to \pi^+\pi^-$, $B_s\to K^+K^-$ 
and $B_d\to\pi^\mp K^\pm$, $B_s\to\pi^\pm K^\mp$. These approaches
are particularly promising from a practical point of view: BaBar, Belle and 
CLEO-III can probe $\gamma$ nicely through $B\to\pi K$ modes, whereas 
the $U$-spin strategies, requiring also the measurement of $B_s$-meson 
decays, are interesting for Tevatron-II and can be fully exploited at 
BTeV and the LHC experiments.

\section{Extraction of $\gamma$ from $B\to\pi K$ Decays}
Let us first point out some interesting features of the $B\to\pi K$
system. Because of the small ratio 
$|V_{us}V_{ub}^\ast/(V_{ts}V_{tb}^\ast)|\approx0.02$, these decays
are dominated by QCD penguin topologies, despite their loop
suppression. Due to the large top-quark mass, we have also to care 
about electroweak (EW) penguins. In the case of $B^+\to\pi^+K^0$ and 
$B^0_d\to\pi^-K^+$, these topologies contribute only in color-suppressed form 
and are hence expected to play a minor r\^ole, whereas they contribute also
in color-allowed form to $B^0_d\to\pi^0K^0$ and $B^+\to\pi^0K^+$ and may 
here even compete with tree-diagram-like topologies.

Using the isospin flavor symmetry of strong interactions, we may 
derive the following amplitude relations: 
\begin{displaymath}
\sqrt{2}A(B^+\to\pi^0K^+)+A(B^+\to\pi^+K^0)=
\sqrt{2}A(B^0_d\to\pi^0K^0)+A(B^0_d\to\pi^-K^+)
\end{displaymath}
\begin{equation}\label{ampl-rel}
=-\biggl[|T+C|e^{i\delta_{T+C}}e^{i\gamma}+
P_{\rm ew}\biggr]\propto\left[e^{i\gamma}+q_{\rm ew}\right],
\end{equation}
where $T$ and $C$ denote the strong amplitudes of color-allowed and
color-suppressed tree-diagram-like topologies, respectively, $P_{\rm ew}$ 
is due to color-allowed and color-suppressed EW penguins, $\delta_{T+C}$ 
is a CP-conserving strong phase, and $q_{\rm ew}$ denotes the ratio of 
the EW to tree-diagram-like topologies. A relation with an analogous 
phase structure holds also for the ``mixed'' $B^+\to\pi^+ K^0$, 
$B_d^0\to\pi^- K^+$ system. Because of these relations, the following
combinations of $B\to\pi K$ decays were considered in the literature to
probe $\gamma$:
\begin{itemize}
\item The ``mixed'' $B^\pm\to\pi^\pm K$, $B_d\to\pi^\mp K^\pm$ 
system \cite{PAPIII}--\cite{defan}.
\item The ``charged'' $B^\pm\to\pi^\pm K$, $B^\pm\to\pi^0K^\pm$ 
system \cite{NR}--\cite{BF-neutral1}.
\item The ``neutral'' $B_d\to\pi^0 K$, $B_d\to\pi^\mp K^\pm$ 
system \cite{BF-neutral1,BF-neutral2}.
\end{itemize}
Interestingly, already CP-averaged $B\to\pi K$ branching ratios
may lead to non-trivial constraints on $\gamma$ \cite{FM,NR}. In order
to {\it determine} this angle, also CP-violating rate differences have
to be measured. To this end, we introduce the following observables
\cite{BF-neutral1}:
\begin{equation}\label{mixed-obs}
\left\{\begin{array}{c}R\\A_0\end{array}\right\}
\equiv\left[\frac{\mbox{BR}(B^0_d\to\pi^-K^+)\pm
\mbox{BR}(\overline{B^0_d}\to\pi^+K^-)}{\mbox{BR}(B^+\to\pi^+K^0)+
\mbox{BR}(B^-\to\pi^-\overline{K^0})}\right]\frac{\tau_{B^+}}{\tau_{B^0_d}}
\end{equation}
\begin{equation}\label{charged-obs}
\left\{\begin{array}{c}R_{\rm c}\\A_0^{\rm c}\end{array}\right\}
\equiv2\left[\frac{\mbox{BR}(B^+\to\pi^0K^+)\pm
\mbox{BR}(B^-\to\pi^0K^-)}{\mbox{BR}(B^+\to\pi^+K^0)+
\mbox{BR}(B^-\to\pi^-\overline{K^0})}\right]
\end{equation}
\begin{equation}\label{neut-obs}
\left\{\begin{array}{c}R_{\rm n}\\A_0^{\rm n}\end{array}\right\}
\equiv\frac{1}{2}\left[\frac{\mbox{BR}(B^0_d\to\pi^-K^+)\pm
\mbox{BR}(\overline{B^0_d}\to\pi^+K^-)}{\mbox{BR}(B^0_d\to\pi^0K^0)+
\mbox{BR}(\overline{B^0_d}\to\pi^0\overline{K^0})}\right],
\end{equation}
where the $R_{\rm (c,n)}$ are ratios of CP-averaged branching ratios and
the $A_0^{\rm (c,n)}$ represent CP-violating observables. 

If we employ the $SU(2)$ flavor symmetry, which implies (\ref{ampl-rel}), 
and make plausible dynamical assumptions, concerning mainly the 
smallness of certain rescattering processes, we obtain parametrizations 
of the following kind \cite{defan,BF-neutral1} (for alternative ones, see 
\cite{neubert}):
\begin{equation}\label{obs-par}
R_{({\rm c,n})},\, A_0^{({\rm c,n})}=
\mbox{functions}\left(q_{({\rm c,n})}, r_{({\rm c,n})},
\delta_{({\rm c,n})}, \gamma\right).
\end{equation}
Here $q_{({\rm c,n})}$ denotes the ratio of EW penguins to ``trees'',
$r_{({\rm c,n})}$ is the ratio of ``trees'' to QCD penguins, and
$\delta_{({\rm c,n})}$ the strong phase between ``trees'' 
and QCD penguins for the mixed, charged and neutral $B\to\pi K$ systems,
respectively. The EW penguin parameters $q_{({\rm c,n})}$
can be fixed through theoretical arguments: in the mixed system 
\cite{PAPIII}--\cite{GR}, we have $q\approx0$, as EW penguins contribute only 
in color-suppressed form; in the charged and neutral $B\to\pi K$ systems,
$q_{\rm c}$ and $q_{\rm n}$ can be fixed through the $SU(3)$ flavor
symmetry without dynamical assumptions \cite{NR}--\cite{BF-neutral2}. 
The $r_{({\rm c,n})}$ can be determined with the help of additional 
experimental information: in the mixed system, $r$ can be fixed 
through arguments based on factorization \cite{QCD-fact,PAPIII,GR} or 
$U$-spin, as we will see below, whereas $r_{\rm c}$ and $r_{\rm n}$ can be 
determined from the CP-averaged $B^\pm\to\pi^\pm\pi^0$ branching ratio
by using only the $SU(3)$ flavor symmetry \cite{GRL,NR}. The uncertainties
arising in this program from $SU(3)$-breaking effects can be reduced
through the QCD factorization approach \cite{QCD-fact,BBNS3}, which is 
moreover in favour of small rescattering processes. For simplicity, we shall
neglect such FSI effects below; more detailed discussions can be found in
\cite{FSI}. 

Since we are in a position to fix the parameters $q_{({\rm c,n})}$ 
and $r_{({\rm c,n})}$, we may determine $\delta_{({\rm c,n})}$ 
and $\gamma$ from the observables given in (\ref{obs-par}). This can 
be done separately for the mixed, charged and neutral $B\to\pi K$ 
systems. It should be emphasized that also CP-violating rate 
differences have to be measured to this end. Using 
just the CP-conserving observables $R_{({\rm c,n})}$, we may obtain 
interesting constraints on $\gamma$. In contrast to $q_{({\rm c,n})}$ 
and $r_{({\rm c,n})}$, the strong phase $\delta_{({\rm c,n})}$ suffers 
from large hadronic uncertainties. However, we can get rid of 
$\delta_{({\rm c,n})}$ by keeping it as a ``free'' variable, yielding 
minimal and maximal values for $R_{({\rm c,n})}$:
\begin{equation}\label{const1}
\left.R^{\rm ext}_{({\rm c,n})}\right|_{\delta_{({\rm c,n})}}=
\mbox{function}\left(q_{({\rm c,n})},r_{({\rm c,n})},\gamma\right).
\end{equation}
Keeping in addition $r_{({\rm c,n})}$ as a free variable, we obtain 
another -- less restrictive -- minimal value for $R_{({\rm c,n})}$:
\begin{equation}\label{const2}
\left.R^{\rm min}_{({\rm c,n})}\right|_{r_{({\rm c,n})},\delta_{({\rm c,n})}}
=\mbox{function}\left(q_{({\rm c,n})},\gamma\right)\sin^2\gamma.
\end{equation}
These extremal values of $R_{({\rm c,n})}$ imply 
constraints on $\gamma$, since the cases corresponding to
$R^{\rm exp}_{({\rm c,n})}< R^{\rm min}_{({\rm c,n})}$
and $R^{\rm exp}_{({\rm c,n})}> R^{\rm max}_{({\rm c,n})}$
are excluded. The present experimental status is summarized in 
Table~\ref{tab1}. We observe that both the CLEO and the Belle
data point towards $R_{\rm c}>1$ and $R_{\rm n}<1$, whereas the
central values of the BaBar collaboration are close to one, with 
a small preferrence of $R_{\rm c}>1$.

\begin{table}
\begin{tabular}{lrrr}
\hline
  & \tablehead{1}{r}{b}{CLEO \cite{CLEO-BpiK}}
  & \tablehead{1}{r}{b}{BaBar \cite{BaBar-BpiK}}
  & \tablehead{1}{r}{b}{Belle \cite{Belle-BpiK}}   \\
\hline
$R$ & $1.00\pm0.30$ & $0.97\pm0.23$ & $1.50\pm0.66$ \\
$R_{\rm c}$ & $1.27\pm0.47$ & $1.19\pm0.35$ & $2.38\pm1.12$ \\
$R_{\rm n}$ & $0.59\pm0.27$ & $1.02\pm0.40$ & $0.60\pm0.29$ \\
\hline
\end{tabular}
\caption{Present experimental status of the observables $R_{({\rm c,n})}$.
For the evaluation of $R$, we have used 
$\tau_{B^+}/\tau_{B^0_d}=1.060\pm0.029$.}
\label{tab1}
\end{table}

\begin{figure}[b]
\centerline{\rotate[r]{
\epsfysize=8.5truecm
{\epsffile{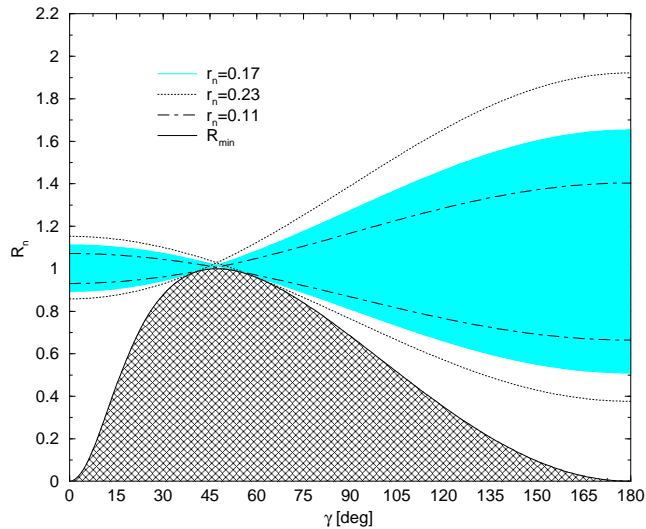}}}}
\caption{The dependence of the extremal values of $R_{\rm n}$ 
on $\gamma$ for $q_{\rm n}=0.68$.}\label{fig1}
\end{figure}

In Fig.\ \ref{fig1}, we show the dependence of (\ref{const1}) and
(\ref{const2}) on $\gamma$ for the neutral $B\to\pi K$ system; the
charged $B\to\pi K$ curves look very similar \cite{BF-neutral2}. Here 
the crossed region below the $R_{\rm min}$ curve, which is described by 
(\ref{const2}), is excluded. On the other hand, the shaded region is the 
allowed range (\ref{const1}) for $R_{\rm n}$, arising in the case of 
$r_{\rm n}=0.17$. This figure allows us to read off immediately the allowed 
range for $\gamma$ for a given value of $R_{\rm n}$. In order to illustrate 
this feature, let us assume that $R_{\rm n}=0.6$ has been measured, which
would be in accordance with the central values of CLEO and Belle in 
Table~\ref{tab1}. In this case, the $R_{\rm min}$ curve implies 
$0^\circ\leq\gamma\leq19^\circ\,\lor\,97^\circ\leq\gamma
\leq180^\circ$. If we use additional information on $r_{\rm n}$, we may 
put even stronger constraints on $\gamma$. For $r_{\rm n}=0.17$, we 
obtain, for instance, the allowed range $134^\circ\leq\gamma\leq180^\circ$. 
It is interesting to note that the $R_{\rm min}$ curve is only effective 
for $R_{\rm n}<1$. Assuming $R_{\rm c}=1.3$ to illustrate implications of
the CP-averaged charged $B\to\pi K$ branching ratios, (\ref{const2}) is 
not effective and $r_{\rm c}$ has to be fixed in order to constrain 
$\gamma$. Using $r_{\rm c}=0.21$, we obtain $84^\circ\leq\gamma\leq180^\circ$.
Although it is too early to draw definite conclusions, it is important to 
emphasize that the second quadrant for $\gamma$, i.e.\ $\gamma\geq 90^\circ$, 
would be preferred in these cases. Interestingly, such a situation would be 
in conflict with the standard analysis of the unitarity triangle 
\cite{UT-fits}, yielding $\gamma\sim 60^\circ$. Here the stringent present
experimental lower bound on $\Delta M_s$ implies $\gamma<90^\circ$. 

Another ``puzzle'' may arise from CP-conserving strong phases, which can 
also be constrained through the observables $R_{({\rm c,n})}$ 
\cite{BF-neutral2}. Interestingly, the CLEO and Belle data point towards 
$\cos\delta_{\rm c}>0$ and $\cos\delta_{\rm n}<0$, whereas we would expect 
equal signs for these quantities. Moreover, $\cos\delta_{\rm n}<0$ would 
be in conflict with factorization. 

Consequently, the present CLEO and Belle data point towards a ``puzzling''
situation, whereas no such discrepancies are indicated by the results of 
the BaBar collaboration. It is of course too early to draw any definite
conclusions. However, if future data should confirm such a picture, it 
may be an indication for new physics or large flavor-symmetry-breaking
effects \cite{BF-neutral2}. Further studies are desirable to distinguish
between these cases. Since $B\to\pi K$ modes are governed by penguin
processes, they actually represent sensitive probes for new 
physics \cite{BpiK-NP}.

Due to the recent theoretical progress in the description of 
$B\to\pi K,\pi\pi$ decays, the theoretical uncertainties of $r_{\rm c,n}$
and $q_{\rm c,n}$ can be reduced to the level of \cite{BBNS3}
\begin{equation}
{\cal O}\left(\frac{1}{N_{\rm C}}\times\frac{m_s-m_d}{\Lambda_{\rm QCD}}
\times\frac{\Lambda_{\rm QCD}}{m_b}\right)={\cal O}\left(\frac{1}{N_{\rm C}}
\times\frac{m_s-m_d}{m_b}\right),
\end{equation}
and confidence into dynamical assumptions related to rescattering effects 
can be gained. Making more extensive use of QCD factorization, approaches 
complementary to the ones discussed above, which rely on a ``minimal'' input 
from theory, are provided. As a first step, we may use that the CP-conserving 
strong phase $\delta_{\rm c}$ is predicted to be very small in QCD 
factorization, so that $\cos\delta_{\rm c}$ governing $R_{\rm c}$ is close 
to one. As a second step, information on $\gamma$ can be obtained from the 
predictions for the branching ratios and the observables $R_{\rm (c,n)}$. 
Finally, the information from all CP-averaged $B\to\pi K,\pi\pi$ branching 
ratios can be combined into a single global fit for the allowed region in
the $\overline{\rho}$--$\overline{\eta}$ plane \cite{BBNS3,lacker}.
For these approaches, it is of course crucial that the corrections
entering in the QCD factorization formalism at the $\Lambda_{\rm QCD}/m_b$
level can be controlled reliably. As argued in a recent paper 
\cite{charm-pengs}, non-perturbative contributions with the same quantum
numbers as penguin topologies with internal charm- and up-quark exchanges 
may play an important r\^ole in this context. The issue of 
$\Lambda_{\rm QCD}/m_b$ corrections in phenomenological analyses will 
certainly continue to be a hot topic in the future.

\section{$U$-Spin Strategies}
Let us now focus on strategies to extract $\gamma$ from pairs of 
$B$-meson decays, which are related to each other through the 
$U$-spin flavor symmetry of strong interactions. In order to
deal with non-leptonic $B$ decays, $U$-spin offers an important
tool, and first approaches to extract CKM phases were already 
pointed out in 1993 \cite{snowmass}. However, the great power of the 
$U$-spin symmtery to determine weak phases and hadronic parameters 
was noticed just recently in the strategies proposed in 
\cite{RF-BdsPsiK}--\cite{skands}. Since these methods involve also 
decays of $B_s$-mesons, $B$ experiments at hadron colliders 
are required to implement them in practice. At Tevatron-II, we 
will have first access to the corresponding modes and interesting 
results are expected \cite{CDF-2}. In the era of BTeV and the LHC, 
the $U$-spin strategies can then be fully exploited \cite{LHC-Report},
as emphasized by Stone at this symposium. In the following discussion, 
we shall focus on two particularly promising approaches, using the
$B_d\to \pi^+\pi^-$, $B_s\to K^+K^-$ \cite{RF-BsKK} and 
$B_d\to\pi^\mp K^\pm$, $B_s\to\pi^\pm K^\mp$ \cite{BspiK} systems.

\subsection{Extraction of $\beta$ and $\gamma$ from  
$B_d\to\pi^+\pi^-$, $B_s\to K^+K^-$ Decays}
Looking at the corresponding Feynman diagrams, we observe that 
$B_s\to K^+K^-$ is obtained from $B_d\to\pi^+\pi^-$ by interchanging 
all down and strange quarks. The structure of the corresponding decay 
amplitudes is given as follows \cite{RF-BsKK}:
\begin{equation}\label{Bdpipi-ampl0}
A(B_d^0\to\pi^+\pi^-)={\cal C}\left[e^{i\gamma}-d e^{i\theta}\right]
\end{equation}
\begin{equation}\label{BsKK-ampl0}
A(B_s^0\to K^+K^-)=\lambda\,{\cal C}'\left[e^{i\gamma}+
\left(\frac{1-\lambda^2}{\lambda^2}\right)
d'e^{i\theta'}\right],
\end{equation}
where ${\cal C}$, ${\cal C}'$ are CP-conserving strong amplitudes, and
$d e^{i\theta}$, $d'e^{i\theta'}$ measure, sloppily speaking, ratios
of penguin to tree amplitudes. Using these general parametrizations, 
we obtain the following expressions for the direct and mixing-induced 
CP asymmetries:
\begin{eqnarray}
{\cal A}_{\rm CP}^{\rm dir}(B_d\to\pi^+\pi^-)&=&
\mbox{function}(d,\theta,\gamma)\label{ASYM-1}\\
{\cal A}_{\rm CP}^{\rm mix}(B_d\to\pi^+\pi^-)&=&
\mbox{function}(d,\theta,\gamma,\phi_d=2\beta)
\end{eqnarray}
\vspace*{-0.9truecm}
\begin{eqnarray}
{\cal A}_{\rm CP}^{\rm dir}(B_s\to K^+K^-)&=&
\mbox{function}(d',\theta',\gamma)\\
{\cal A}_{\rm CP}^{\rm mix}(B_s\to K^+K^-)&=&
\mbox{function}(d',\theta',\gamma,\phi_s\approx0).\label{ASYM-4}
\end{eqnarray}
Consequently, we have four observables, depending on six ``unknowns''. 
However, since $B_d\to\pi^+\pi^-$ and $B_s\to K^+K^-$ are related to each 
other by interchanging all down and strange quarks, the $U$-spin flavor 
symmetry of strong interactions implies
\begin{equation}\label{U-spin-rel}
d'e^{i\theta'}=d\,e^{i\theta}.
\end{equation}
Using this relation, the four observables (\ref{ASYM-1})--(\ref{ASYM-4}) 
depend on the four quantities $d$, $\theta$, $\phi_d=2\beta$ and $\gamma$, 
which can hence be determined. It should be emphasized that no dynamical
assumptions about rescattering processes have to be made in this approach,
which is an important conceptual advantage in comparison with the $B\to\pi K$
strategies discussed above. The theoretical accuracy is hence only limited
by $U$-spin-breaking effects. Theoretical considerations allow us to 
gain confidence into (\ref{U-spin-rel}), which does not receive 
$U$-spin-breaking corrections in factorization \cite{RF-BsKK}. Moreover,
there are general relations between observables of $U$-spin-related decays, 
allowing experimental insights into $U$-spin breaking 
\cite{RF-BdsPsiK,RF-BsKK,RF-VV,gronau-U-spin}.

The $U$-spin arguments can be minimized, if we employ the 
$B^0_d$--$\overline{B^0_d}$ mixing phase $\phi_d=2\beta$ as an input, 
which can be determined straightforwardly through $B_d\to J/\psi K_{\rm S}$. 
The observables
${\cal A}_{\rm CP}^{\rm dir}(B_d\to\pi^+\pi^-)$ and
${\cal A}_{\rm CP}^{\rm mix}(B_d\to\pi^+\pi^-)$ allow us then to 
eliminate the strong phase $\theta$ and to determine $d$ as a function of
$\gamma$. Analogously, ${\cal A}_{\rm CP}^{\rm dir}(B_s\to K^+K^-)$ and 
${\cal A}_{\rm CP}^{\rm mix}(B_s\to K^+K^-)$ allow us to eliminate 
the strong phase $\theta'$ and to determine $d'$ as a function of
$\gamma$. The corresponding contours in the $\gamma$--$d$
and $\gamma$--$d'$ planes can be fixed in a {\it theoretically clean} way.
Using now the $U$-spin relation $d'=d$, these contours allow the 
determination both of the CKM angle $\gamma$ and of the hadronic quantities 
$d$, $\theta$, $\theta'$; for a detailed illustration, see \cite{RF-BsKK}.

This approach is very promising for Tevatron-II and 
the LHC era, where experimental accuracies for $\gamma$ of 
${\cal O}(10^\circ)$ \cite{CDF-2} and ${\cal O}(1^\circ)$ \cite{LHC-Report}
may be achieved, respectively. It should be emphasized that not only
$\gamma$, but also the hadronic parameters $d$, $\theta$, $\theta'$ are
of particular interest, as they can be compared with theoretical 
predictions, thereby allowing valuable insights into hadron dynamics. 
For strategies to probe $\gamma$ and constrain hadronic penguin 
parameters using a variant of the $B_d\to \pi^+\pi^-$, $B_s\to K^+K^-$ 
approach, where the latter decay is replaced through $B_d\to\pi^\mp K^\pm$, 
the reader is referred to \cite{U-variant}.

\subsection{Extraction of $\gamma$ from $B_{(s)}\to \pi K$ Decays}
Another interesting pair of decays, which are related to each other by
interchanging all down and strange quarks, is the $B^0_d\to\pi^-K^+$, 
$B^0_s\to\pi^+K^-$ system \cite{BspiK}. In the strict $U$-spin limit, 
the corresponding decay amplitudes can be parametrized as follows:
\begin{equation}\label{PAR1}
A(B^0_d\to\pi^-K^+)=-P\left(1-re^{i\delta}e^{i\gamma}\right)
\end{equation}
\begin{equation}\label{PAR2}
A(B^0_s\to\pi^+K^-)=P\sqrt{\epsilon}
\left(1+\frac{1}{\epsilon}\, r e^{i\delta}e^{i\gamma}\right),
\end{equation}
where $P$ denotes a CP-conserving complex amplitude, 
$\epsilon\equiv\lambda^2/(1-\lambda^2)$, $r$ is a real parameter, and
$\delta$ a CP-conserving strong phase. At first sight, it appears as 
if $\gamma$, $r$ and $\delta$ could be determined from the ratio of the 
CP-averaged rates and the two CP asymmetries provided by these 
modes.\footnote{Note that these observables are independent of $P$.} 
However, because of the relation 
\begin{displaymath}
|A(B^0_d\to\pi^-K^+)|^2-|A(\overline{B^0_d}\to\pi^+K^-)|^2=4 r \sin\delta
\sin\gamma 
\end{displaymath}
\begin{equation}\label{RU-rel}
=-\left[|A(B^0_s\to\pi^+K^-)|^2-|A(\overline{B^0_s}\to\pi^-K^+)
|^2\right],
\end{equation}
we have actually only two independent observables, so that the three
parameters $\gamma$, $r$ and $\delta$ cannot be determined. To this
end, the overall normalization $P$ has to be fixed, requiring a further
input. Assuming that rescattering processes play a minor r\^ole 
and that color-suppressed EW penguins can be neglected as well, the 
isospin symmetry implies
\begin{equation}\label{PAR3}
P=A(B^+\to\pi^+K^0).
\end{equation}
In order to extract $\gamma$ and the hadronic parameters, it is useful
to introduce observables $R_s$ and $A_s$ by replacing $B_d\to\pi^\mp K^\pm$
through $B_s\to\pi^\pm K^\mp$ in (\ref{mixed-obs}). Using (\ref{PAR1}), 
(\ref{PAR2}) and (\ref{PAR3}), we then obtain 
\begin{equation}
R_s=\epsilon+2r\cos\delta\cos\gamma+\frac{r^2}{\epsilon}
\end{equation}
\begin{equation}
A_s=-2r\sin\delta\sin\gamma=-A_0.
\end{equation}
Together with the parametrization for $R$ as sketched in (\ref{obs-par}), 
these observables allow the determination of all relevant parameters. 
The extraction of $\gamma$ and $\delta$ is analogous to the ``mixed'' 
$B_d\to\pi^\mp K^\pm$, $B^\pm\to\pi^\pm K$ approach discussed above. 
However, now the advantage is that the $U$-spin counterparts 
$B_s\to\pi^\pm K^\mp$ of $B_d\to\pi^\mp K^\pm$ allow us 
to determine also the parameter $r$ without using arguments related to 
factorization
\cite{BspiK}:
\begin{equation}
r=\sqrt{\epsilon\left[\frac{R+R_s-1-\epsilon}{1+\epsilon}\right]}.
\end{equation}
On the other hand, we still have to make dynamical assumptions concerning 
rescattering and color-suppressed EW penguin effects. The theoretical
accuracy is further limited by $SU(3)$-breaking effects. An interesting
consistency check is provided by the relation $A_s=-A_0$, which is due 
to (\ref{RU-rel}). A variant of this approach using the CKM angle $\beta$ 
as an additional input was proposed in \cite{CW-BspiK}.

\section{Conclusions}
There are many strategies to determine $\gamma$, suffering unfortunately 
in several cases from experimental problems. The approaches discussed 
above, employing penguin processes, are on the other hand very promising 
from a practical point of view and exhibit further interesting features. 
As a by-product, they also allow us to determine strong phases and other
hadronic parameters, allowing comparisons with theoretical predictions.
Moreover, these strategies are sensitive probes for the physics lying 
beyond the Standard Model, which may lead to discrepancies in the extraction 
of $\gamma$ or the hadronic quantities. Let us hope that signals for new
physics will actually emerge this way.

\vfill\eject

\end{document}